\newif\ifarxiv
\arxivtrue                 % article format
\ifarxiv
  \documentclass[11pt]{article}
  \usepackage[margin=1in]{geometry}
  \usepackage{graphicx}
\else
  \documentclass{iopjournal} % use your existing IOP class/options
\fi

\usepackage{amsmath}
\usepackage{url}
 
\usepackage[numbers,sort&compress]{natbib}
\usepackage{doi}
\usepackage{hyperref}
\usepackage{booktabs,longtable,siunitx}

\hypersetup{
    colorlinks=true,
    citecolor=blue,
    urlcolor=blue
}

\providecommand{\ack}[1]{%
  \section*{Acknowledgments}#1\par}

\providecommand{\roles}[1]{%
  \section*{Author contributions}#1\par}

\providecommand{\funding}[1]{%
  \section*{Funding}#1\par}

\providecommand{\data}[1]{%
  \section*{Data availability}#1\par}

\providecommand{\suppdata}[1]{%
  \section*{Supplementary data}#1\par}

\begin{document}

\ifarxiv
    \title{Electron Kinetics in the Sheath of Atmospheric Pressure
    Plasmas in Helium and Helium:H$_2$O Vapor}

    \author{Uwe Kortshagen\\
    Department of Mechanical Engineering\\
    University of Minnesota, Minneapolis, MN 55455, USA\\
    \texttt{kortshagen@umn.edu}}

  \date{} % Suppress the automatically printed date
  \maketitle

\else
    \articletype{Paper} %	 e.g. Paper, Letter, Topical Review...
    
    \title{Electron Kinetics in the Sheath of 
     Atmospheric Pressure Plasmas in Helium and Helium:H$_2$O Vapor}

    \author{Uwe Kortshagen$^1,*$\orcid{0000-0001-5944-3656}}

    \affil{$^1$Department of Mechanical Engineering, University of Minnesota, Minneapolis, USA}

    \affil{$^*$Author to whom any correspondence should be addressed.}

    \email{kortshagen@umn.edu}


\fi

\begin{abstract}
The electron energy distribution function (EEDF) plays a central role in determining the rates of electron-induced reactions in the bulk plasma, the sheath, and at the plasma-surface interface. However, the electron kinetics within the sheath and near the surface are generally not well described by conventional Boltzmann solvers, which neglect the sheath electric field and electron losses to the surface. In this study, we investigate electron kinetics in atmospheric pressure plasmas in pure helium and helium-water vapor mixtures using a hybrid model that combines a fluid description of the collisional sheath with a particle kinetic Monte Carlo model for electrons. The floating wall potential is determined taking into account electron diffusion through the sheath and the non-Maxwellian EEDF. Our results demonstrate that electron kinetics within the sheath can exhibit both local and nonlocal behavior, even at atmospheric pressure. The transition between these regimes is governed primarily by the ratio of the electron energy relaxation length to the sheath thickness. At high plasma densities, where the sheath thickness becomes smaller than the electron energy relaxation length, the EEDFs approach the nonlocal limit, exhibiting significant changes in electron kinetic energy toward the wall. The addition of water vapor reduces the energy relaxation length and shifts the electron kinetics toward the local regime. In the nonlocal regime, the electron mean kinetic energy decreases substantially toward the wall, accompanied by reductions of more than an order of magnitude in electron-impact excitation and ionization rate coefficients. These findings demonstrate that accurate modeling of electron-induced processes in the sheath and at plasma-surface interfaces requires accounting for nonlocal electron kinetics, sheath electric fields, and electron losses to the surface, even at atmospheric pressure.
\end{abstract}

\section{Introduction}

In low-temperature plasmas, the importance of knowing the electron energy distribution function (EEDF) is well understood. The EEDF determines the rates of all electron-induced chemical reactions in the plasma \cite{liebermanPrinciplesPlasmaDischarges2005,Raizer1991}. Due to the pronounced non-equilibrium in low-temperature plasmas, the EEDF often strongly deviates from a Maxwellian distribution \cite{Raizer1991,golantFundamentalsPlasmaPhysics1980} and needs to be determined using appropriate tools. These tools include Boltzmann equation solvers, such as BOLSIG+\cite{hagelaarSolvingBoltzmannEquation2005} or LoKI-B \cite{tejero-del-cazLisbOnKIneticsBoltzmann2019}, or fully kinetic particle-in-cell calculations \cite{birdsallParticleincellChargedparticleSimulations1991,vahediCapacitiveRFDischarges1993,donkoParticleSimulationMethods2011}.

The vast majority of applications of low-temperature plasmas include the interaction of a plasma with a solid or liquid surface. One example is plasma surface interactions in semiconductor plasma processing \cite{gravesScienceChallengesResearch2024}. Other examples include plasma liquid interactions \cite{bruggemanNonthermalPlasmasContact2009,bruggemanPlasmaLiquidInteractions2016}, for example, for the decontamination of water \cite{vonwoedtkePlasmaLiquidInteractionsChemistry2012}, or for the synthesis of nanomaterials \cite{chenReviewPlasmaLiquid2015a}. The rates of all electron-induced plasma surface interactions depend on the EEDF right at the interface between the plasma and the solid or liquid. An example of this is the rate of electron-induced decomposition of PFAS molecules at the surface of water exposed to a plasma \cite{singhBreakdownProductsPerfluorinated2019,guoTheoreticalExperimentalInsights2022}. Other examples include the interaction of living tissue with plasma in plasma medical applications \cite{vonwoedtkePlasmasMedicine2013}, the dissociation of water by “hot electrons” \cite{jonesExcitedStateNonadiabatic2023,chipmanDissociativeElectronAttachment2007,yueElectronDensityElectron2022}, the formation of nanomaterials at the plasma-liquid interface through the reduction of metal ions by injected plasma electrons \cite{leePreparationNonaggregatedSilver2013, xuSilverNanoparticleSynthesis2023}, and the charging of nanoparticles in nano-dusty plasmas \cite{gattiAnalyticalModelParticle2008, khrapakParticleChargeBulk2005}. 

Given that plasma surface interactions are crucial for many low-temperature plasma applications and that electrons play a central role, it is surprising that more attention has not been paid to the EEDF in the sheath and at the plasma surface interface. One may argue that, in particle-in-cell simulations, we can find the EEDF with sufficient accuracy; yet, to date, many multi-scale plasma models are based on fluid or hybrid simulations that rely on rate coefficients often determined by traditional Boltzmann solvers \cite{liuFluidModelPlasma2021,kenileyMultiphaseModelingDC2022,parseyGuidedPlasmaJets2021}.  However, conventional spatially homogeneous or local Boltzmann calculations, without an appropriate representation of the sheath electric field and electron–surface boundary conditions \cite{willertMicrophysicalBoundaryCondition2026} may not reliably predict the EEDF within the sheath.

In this paper, we argue that the EEDF within the sheath and at the plasma surface interface, here also referred to as the wall, is determined by the ratio of the plasma sheath length, $L_s$, which is on the order of a few Debye lengths, and the electron energy relaxation length, $\lambda_\varepsilon$, \cite{bernsteinElectronEnergyDistributions1954,tsendinEnergyDistributionElectrons1974,kolobovNonlocalElectronKinetics1995,kortshagenElectronIonDistribution1995} which describes the distance over which electrons lose a significant fraction of their kinetic energy. The central hypothesis of this paper is that the application of traditional, local Boltzmann solvers within the sheath and at the plasma-surface interface may be incorrect even at pressures as high as atmospheric, and that a more detailed description that correctly accounts for sheath electric fields and the presence of the wall needs to be developed. 

\section{Background}

The space in front of any surface bounding the plasma is dominated by the sheath, as has been known for over a century \cite{langmuirPositiveIonCurrents}. It is fair to say that the vast majority of literature on plasma sheaths assumes a Maxwellian EEDF
\cite{selfExactSolutionCollisionless1963,kochEffectsCollisionsPlasma1989,godyakSmoothPlasmasheathTransition1990,k_-u_riemann_bohm_1991}. However, this may not be a good assumption, because (i) the EEDF in the bulk plasma often strongly deviates from a Maxwellian, and (ii) electron losses to the surface should modify the EEDF close to the wall, as also pointed out in the recent perspective paper by Kolobov and Godyak \cite{kolobovElectronKineticsLowtemperature2019}.

The concept of a non-Maxwellian EEDF within the sheath, of course, is inherent in the idea of measuring the EEDF with a Langmuir probe. The Druyvesteyn method, introduced in 1930 \cite{druyvesteynNiedervoltbogen1930}, is a method to determine the EEDF, Maxwellian or non-Maxwellian, from the second derivative of a Langmuir probe characteristic. However, Druyvesteyn’s original derivation assumed that electrons do not perform collisions in the probe sheath and that the EEDF is spherically symmetric \cite{jeallenApplicabilityDruyvesteynMethod1978, godyakProbeMeasurementsElectronenergy2011}. Moreover, the derivation neglected any details about the sheath itself, such as its potential distribution.

Kushner recognized that the electron density variation within the sheath should be described differently from using the conventional Boltzmann factor in the case of a non-Maxwellian EEDF at the sheath boundary \cite{kushnerFloatingSheathPotentials1985}. He solved the sheath problem for a ``two-temperature Maxwellian" and showed a clear influence of the non-Maxwellian EEDF on the sheath potential. However, in his discussion, Kushner assumed that the sheath was collisionless, which we argue is an unnecessarily restrictive assumption. 

Arslanbekov et al. developed a probe theory valid at intermediate and high pressures \cite{arslanbekovProbeMeasurementsElectron1994}. The authors derived an expression linking the EEDF at the probe surface to that in the bulk plasma. However, they did not consider how electron loss to the probe would modify the EEDF at the sheath boundary. While this may be permissible for the cylindrical and spherical probes that they considered, as the influence of the probe on the plasma ideally is negligible, it may be a poor assumption for a planar surface in contact with the plasma, as is the case in most plasma surface interactions. Equally, Jauberteau and Jauberteau~\cite{jauberteauLangmuirProbeMeasurements2026} emphasized the importance of accounting for the electron diffusion through the sheath in interpreting probe measurements. 

Bourdon, Chabert, and coworkers used PIC simulations for a fully kinetic simulation of a collisionless sheath and found that the EEDF is strongly affected by wall losses at energies above the wall potential energy \cite{tavantNonisothermalSheathModel2019}. The authors found that for electrons below the wall potential energy, the EEDF was close to a Maxwellian with $k_B T_e$=5 eV, but above the wall potential energy at about 7 eV, the EEDF was rapidly depleted of electrons due to electron wall loss. The authors also found that the mean kinetic energy of the EEDF strongly decreased towards the wall as a function of wall losses. The authors used a kinetic model based on the collisionless Vlasov equation to interpret the findings of their PIC simulations. 

Similarly, Baalrud and co-workers \cite{baalrudInteractionBiasedElectrodes2020}, using PIC simulations and analytical fluid modeling, realized that the presence of the wall introduces “non-isothermal” effects, assuming that an initial Maxwellian distribution in the plasma is modified by wall losses to become non-Maxwellian. They found that the electron energy decreases rapidly as the electrons approach the wall. The authors also identified the loss of high-energy electrons to the wall as the mechanism leading to this rapid drop in electron energy. 

None of these studies addressed the EEDF in a highly collisional sheath at atmospheric pressure, which is the focus of this paper.

\section{Methods and Numerical Model}

To investigate the electron kinetics within the plasma sheath, we developed a coupled hybrid fluid-kinetic framework. The steady-state electrostatic fields and ion dynamics are described by a one-dimensional fluid sheath model, while the electron kinetics are resolved via a particle kinetic Monte Carlo Collision (MCC) code. The two modules are coupled by accounting for electron transport through the sheath and the non-Maxwellian character of the EEDF to determine the position of the floating wall boundary that satisfies the equality of electron and ion fluxes, $\Gamma_e = \Gamma_i$. The modules and their coupling are discussed below. 

\begin{figure}
 \centering
        \includegraphics[width=0.5\textwidth]{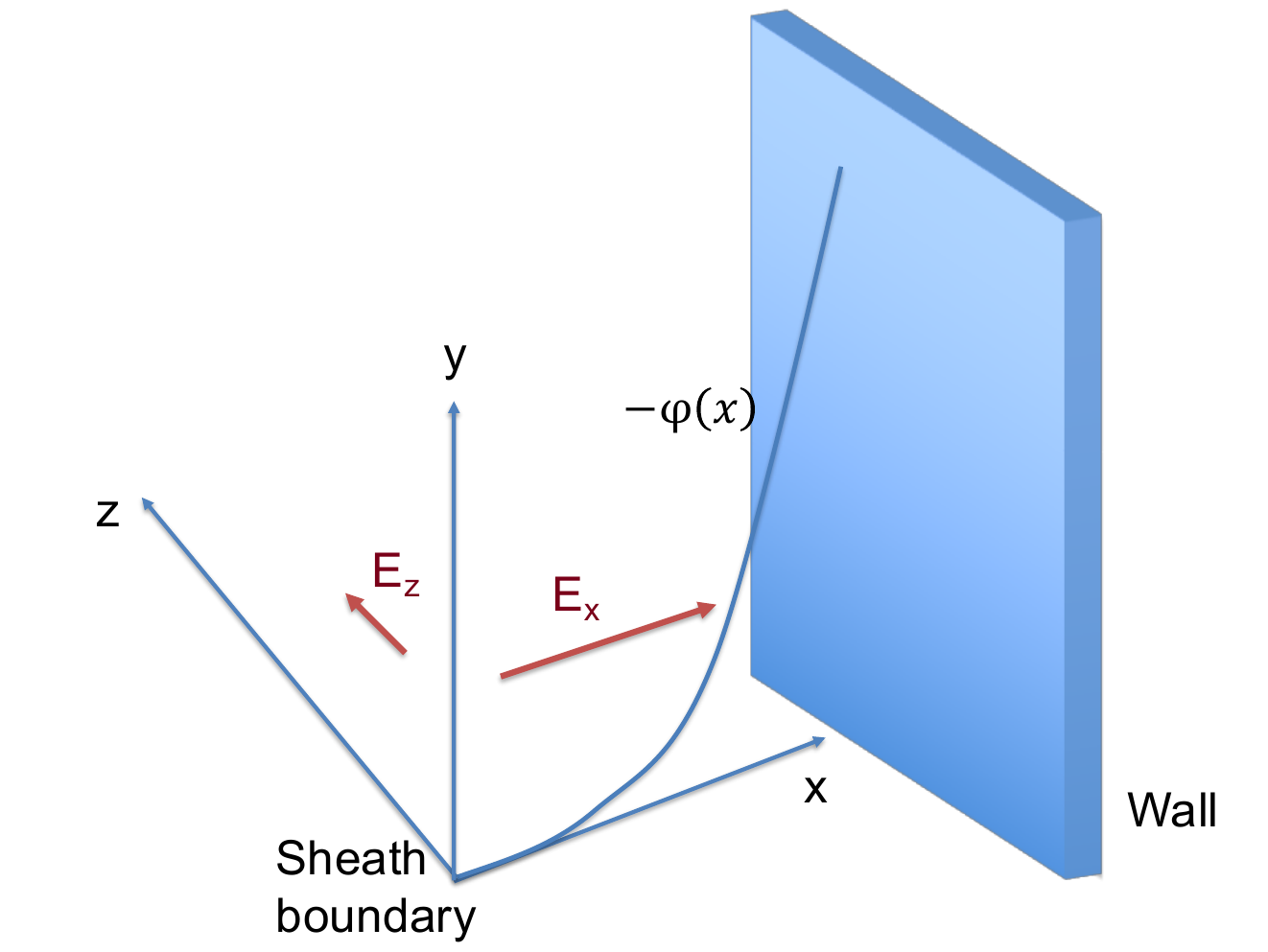}
 \caption{Schematic of the simulation configuration assumed in this study. Electrons are heated by an electric field $E_z$, which is assumed constant throughout the simulation domain. The sheath electric field is denoted by $ E_x$.}
\label{fig:fig1}
\end{figure}

The simulation domain set-up is illustrated in Figure \ref{fig:fig1}. Here we assume a spatially constant, electron-heating electric field $E_z$ parallel to the wall. This is representative, for instance, of the axial electric field in a positive column or the azimuthal electric field in an inductively coupled discharge. It is assumed that outside of the sheath, in the bulk plasma, electrons are heated only by $E_z$. The sheath extends in the positive $x$-direction, with the sheath electric field strongly increasing from the sheath boundary towards the wall. The sheath boundary is designated as $x=0$, the wall is at $x=L_s$, with $L_s$ the sheath thickness.

\subsection{Fluid Sheath Model}

The spatial distribution of the electrostatic potential, $\phi(x)$, and the electric field, $E_x(x)$, within the sheath is determined using a fluid model based on a formalism proposed by Kono \cite{konoStructureCollisionalCollisionless2004}. For the electron kinetic solver, it is convenient to define $\phi(x=0)=\phi_{sb}=0$ at the sheath boundary. However, as described below, Kono's model uses a different reference point for the sheath potential; hence, we introduce a second potential variable, $\psi$, which is referenced to Kono's reference point described below. $\phi$ and $\psi$ are connected by $\phi(x)=\psi(x)-\psi_{sb}$.  

Following Kono, we introduce the standard dimensionless variables normalized by reference electron properties: 
$\Psi = \frac{e\psi}{k_B T_e} $, $\mathcal{E} = \frac{e E_x \lambda_D}{k_B T_e}$, $\mathcal{U} = \frac{v_i}{v_B}$, and $x' = \frac{x}{\lambda_D}$, where $e$ is the elementary charge, $\psi$ the sheath potential, $v_i$ is the directed ion velocity, $v_B = \sqrt{k_B T_e / M_i}$ is the Bohm velocity for an ion of mass $M_i$, and $\lambda_D = \sqrt{\epsilon_0 k_B T_e / n_{e,ref} e^2}$. Ionization in the sheath is neglected; hence, the ion flux $J_i$ is continuous. $n_{e,ref}$ is the electron density at a reference point defined by $J_i = n_{e,ref}v_B$, which also serves as reference for the potential, $\Psi=0$. It is important to note that $n_{e,ref}$ is not the electron density at the sheath boundary, but it can be much smaller than that, because in a highly collisional sheath, the ions may not reach the Bohm velocity anywhere within the sheath. The electron temperature, $T_e$, is defined as $k_B T_e = 2/3<u_{bulk}>$, with $<u_{bulk}>$ the mean kinetic energy of the EEDF in the bulk plasma, as discussed below. 

Assuming cold ions, i.e., neglecting the ion pressure, the Poisson equation and ion momentum equation are reduced to a set of coupled ordinary differential equations in normalized variables:

\begin{align}
  \frac{d\Psi}{dx'} &= -\mathcal{E}, \label{eq:E_field} \\[1ex]
  \frac{d\mathcal{E}}{dx'}    &= \frac{1}{\mathcal{U}} - \exp(\Psi), \label{eq:Poisson} \\[1ex]
  \frac{d\mathcal{U}}{dx'}    &= \frac{\mathcal{E}}{\mathcal{U}} - \nu_m, \label{eq:ion_momentum}
\end{align}

where $\nu_m = \nu_{in}\lambda_D / v_B$ is a normalized ion-neutral momentum exchange collision frequency, and $\nu_{in}$ is the unscaled ion-neutral collision frequency, which we assume to vary linearly with pressure $p$. Equations \ref{eq:E_field} and \ref{eq:Poisson} are the Poisson equation reduced to two first-order ODEs. The term $1/\mathcal{U}$ represents the normalized ion density, assuming ion flux continuity in the sheath and presheath region. $exp(\Psi)$ is the Boltzmann factor for the electron density. While the application of the Boltzmann factor is not strictly correct for a non-Maxwellian EEDF, we here still use it as a reasonable approximation because the low-energy part of the EEDF, which comprises the vast majority of electrons, is often reasonably described by a Maxwellian, with deviations occurring at higher energies. In that sense, our treatment is not fully self-consistent, because the non-Maxwellian EEDF found from the electron simulation below is not fed back into the sheath solution. However, for the placement of the wall boundary, we are accounting for electron diffusion through the sheath and the non-Maxwellian EEDF as described below. 

For the case of a constant ion-neutral collision frequency, which we adopt here, Kono derives analytical asymptotic sheath and pre-sheath solutions by introducing further dimensionless variables $y = \nu_m^{1/3} x'$ and $\tilde\Psi = \Psi - ln(\nu_m^{2/3})$. In the bulk plasma, at sufficiently negative $y$, Kono shows that $\tilde\Psi$ asymptotically approaches $\ln(-y)$. We are using 
$y_{init}=-5$, $\tilde\Psi_{init}=\ln(5)$, and $x'_{init}=y_{init}/\nu_m^{1/3}$. With this, the initial conditions for the integration of equations \ref{eq:E_field}-\ref{eq:ion_momentum} become:
\begin{align}
  \Psi_{init} &= \tilde\Psi_{init} + \ln(\nu_m^{2/3}), \label{eq:phi_zero} \\[1ex]
  \mathcal{E}_{init}    &= -\nu_m^{1/3} \frac{d\Psi}{dy} \Big|_{y_{init}}, \label{eq:e_zero} \\[1ex]
  \mathcal{U}_{init}    &= \frac{\mathcal{E}_{init}}{\nu_m}. \label{eq:u_zero}
\end{align}

For all simulations, we assume that He$^+$ is the dominant ion and a constant ion-neutral collision frequency:
$\nu_{in} = 1.3\times10^7
    \left(
        \frac{p}{100\ \mathrm{Pa}}
    \right)
    \mathrm{s}^{-1}$.

The definition of a sheath boundary in a highly collisional sheath at elevated pressures is somewhat arbitrary, because even the presheath requires a significant electric field, and thus notable space charge, in order to produce the required ion drift \cite{konoStructureCollisionalCollisionless2004}. For instance, in Kono's treatment of the constant ion-neutral collision frequency sheath, the electron density is already by a factor of 2 smaller than the ion density at the sheath boundary. Instead of Kono's choice of the sheath boundary, we adopt the criterion derived by Godyak and Sternberg \cite{godyakSmoothPlasmasheathTransition1990}

\begin{equation}
  \mathcal{U}_{sb} = \frac{v_i}{v_B} = \left(1.0 + \frac{\pi \lambda_D}{2 \lambda_{in}} \right)^{-1/2},
  \label{eq:u_s_velocity}
\end{equation}

with $\lambda_{in}$ the ion neutral mean free path, which yields a definition of the sheath boundary that is closer to quasineutral, particularly for less collisional sheaths. However, different from Godyak and Sternberg, we define $\lambda_D$ using $n_{e,ref}$ at the reference point and not at the sheath boundary, which also yields an agreeable solution. We do not expect the precise choice of the sheath boundary to substantially affect the main conclusions of this study, whose primary focus is the electron kinetics.  

Equations \ref{eq:E_field}-\ref{eq:ion_momentum} are integrated forward in space subject to initial conditions \ref{eq:phi_zero}-\ref{eq:u_zero} using an adaptive-step explicit Runge-Kutta method. The adaptive choice of the wall boundary is discussed below. 

\subsection{Kinetic Electron Monte Carlo Model}

We study electron kinetics using a three-dimensional velocity ($3\text{V}$) and one-dimensional coordinate space ($1\text{D}$) particle kinetic Monte Carlo Collision routine. Individual electron trajectories are tracked continuously in the electric field $\mathbf{E}(x) = E_x(x)\hat{x} + E_z \hat{z}$, where $\hat{x}$ and $\hat{z}$ are unit vectors in $x$ and $z$-directions. The position-dependent sheath electric field $E_x$ is determined from the fluid model described above.

Between collisions, electrons move ballistically. The time between collisions, $\Delta t_{coll}$, is determined from the maximum collision frequency, $\nu_{max}$, derived from the null-collision method \cite{skullerud_stochastic_1968, donkoParticleSimulationMethods2011}:

\begin{equation}
    \Delta t_{\text{coll}} = -\frac{\ln(R)}{\nu_{\text{max}}},
    \label{eq:delta_t_coll}
\end{equation}
where $R \in (0, 1]$ is a uniformly distributed pseudo-random number, and $\nu_{\text{max}}$ is the maximum composite collision frequency evaluated across the entire energy grid, which typically was chosen from 0 to 50 eV. 

Between collisions, electrons are propagated using a second-order velocity Verlet algorithm. At each sub-step $k$, the electron position $x$ and velocity component $v_x$ are updated according to:
\begin{align}
    x_{k+1} &= x_k + v_{x,k} \Delta t + \frac{1}{2} a_{x,k} \Delta t^2, \\
    v_{x,k+1} &= v_{x,k} + \frac{1}{2} \left( a_{x,k} + a_{x,k+1} \right) \Delta t,
\end{align}
where the acceleration $a_{x,k} = -\frac{e}{m_e} E_x(x_k)$ is evaluated via spatial interpolation of the discrete grid for $E_x$ with a cell resolution $\Delta x$. The axial velocity component $v_z$ is updated independently via uniform acceleration $a_z = -\frac{e}{m_e} E_z$. We considered 25 spatial cells, such that $\Delta x = L_s/25$.

To ensure stability and prevent particles from unphysically bypassing multiple cell boundaries between collisions, the time step is constrained by a Courant-Friedrichs-Lewy (CFL) criterion: $\Delta t_{CFL} = \frac{\Delta x}{v_{max}}$, where $v_{max} = 4 \times 10^{6} m/s$ is assumed. When the collision time $\Delta t_{coll} < \Delta t_{CFL}$, we set the integration time step $\Delta t = \Delta t_{coll}$. For $\Delta t_{coll} > \Delta t_{CFL}$, a number of sub-steps, $N_{step}$, was used so that $\Delta t = \Delta t_{coll}/N_{step} < \Delta t_{CFL}$.

After each collision time $\Delta t_{coll}$, electron collisions were evaluated. Electrons either perform a real collision, described by a collision frequency $\nu_{real}$, or a null-collision described with $\nu_{null}$, with $\nu_{real}+\nu_{null} = \nu_{max}$ at all energies. The type of the collision event (e.g., elastic, excitation, ionization, attachment, or null) is chosen based on the cross-section values at a specific energy by another uniformly distributed random number. In the case of a null-collision, the electron velocity vector remains unchanged. Upon a real collision event, the electron energy loss is calculated based on the collision type, and isotropic post-collision scattering is applied.

Boundary conditions are evaluated continuously at each sub-step: trajectories crossing $x < 0$ trigger an electron reinjection, as discussed below. Trajectories exceeding the domain boundary $x \ge L_s$ terminate at the wall, also causing a reinjection.

The electron velocity distribution function (EVDF), $\mathcal{F}(v_x,v_y,v_z)$, is sampled after each integration time step $\Delta t$ and within each of the 25 spatial cells. The EVDF is computed as a velocity histogram created by adding $\Delta t$ to the appropriate bin of the 3D velocity distribution, evaluated at the middle of the time step $\Delta t$. Typically, 101 cells in each of the 3 velocity directions were used for $\sim 10^6$ bins in 3D velocity space, at 25 spatial positions. For a trajectory that remained inside the sheath, the accumulated residence time was equal to the prescribed ballistic flight time. For trajectories crossing either $x=0$ or $x=L_s$, only the fraction of the time step spent inside the simulation domain was recorded. The EEDF, $\mathcal{F}(u)$, with $u$ the electron kinetic energy in eV, is easily derived from the EVDF by summing over all velocity bins corresponding to a kinetic energy within a kinetic energy bin.    

Cross section data were retrieved from LXCat \cite{pitchfordLXCatOpenAccessWebBased2017}. Helium cross sections used the Phelps database \cite{PhelpsDatabaseLXCat,crompton_momentum_1967,crompton_momentum_1970,rappTotalCrossSections1965}, H$_2$O cross sections utilized the Triniti database \cite{TrinitiDatabaseLXCat,yousfi_electronmolecule_1987}. Simulations were run until 10,000 electrons reached the wall. This typically required injecting several million electrons and up to more than 20 million electrons for thick sheaths. 

\subsection{Coupling Electron Kinetic Solver and Sheath Model}

The electron kinetic and fluid sheath models are coupled through the EVDF in the bulk plasma and through the equality of the electron and ion fluxes at the floating wall. The coupling is carried out in three steps: First, the electron distribution in the bulk plasma, i.e., assuming $E_x = 0$, is calculated. This distribution is then used to inject electrons into the sheath. Finally, the wall potential is varied by adjusting the wall position, i.e. the sheath thickness $L_s$, until the electron flux reaching the wall, determined by pilot electron kinetic simulations, agrees with the ion flux obtained from the fluid model.

To determine the bulk EVDF, an electron is followed in a spatially homogeneous plasma in which the only electric field is the constant electron-heating field $E_z$. The same null-collision procedure and collision cross sections as for the full spatially dependent electron simulation are used, and a bulk EVDF is derived from the velocity sampling described above. The electron is initially drawn from an isotropic Maxwellian distribution with a specified initial temperature. It is subsequently accelerated by $E_z$ and undergoes the full range of collisions, including elastic, excitation, ionization, and attachment. The calculation is continued for $10^{-5}$~s, which is sufficiently long for the velocity distribution to become steady with low noise. The effective bulk electron temperature, $T_e$, as defined above, is used to calculate the Bohm velocity and the Debye length used in the sheath fluid simulation.

The velocity distribution of electrons entering the sheath is different from the bulk EVDF, $\mathcal{F}_b(\mathbf{v})$, because faster electrons cross the sheath boundary more frequently. Electrons are therefore sampled from the flux-weighted distribution

\begin{equation}
    P_{\mathrm{inj}}(\mathbf{v}) =
    \frac{
        v_x \mathcal{F}_b(\mathbf{v})
    }{
        \displaystyle
        \int_{v_x>0} v_x \mathcal{F}_b(\mathbf{v})\,d^3v
    },
    \qquad v_x>0.
    \label{eq:flux_weighted_injection}
\end{equation}

A cumulative distribution is constructed from Eq.~\ref{eq:flux_weighted_injection} on the discrete velocity grid. All three velocity components of an injected electron are sampled jointly from this distribution. This procedure retains any bulk EVDF anisotropy produced by $E_z$.

The absolute electron flux entering the sheath is obtained from the positive-$v_x$ moment of the bulk EVDF,

\begin{equation}
    \left\langle v_x^+ \right\rangle =
    \frac{
        \displaystyle
        \int_{v_x>0} v_x \mathcal{F}_b(\mathbf{v})\,d^3v
    }{
        \displaystyle
        \int \mathcal{F}_b(\mathbf{v})\,d^3v
    }.
    \label{eq:positive_vx_moment}
\end{equation}

The electron flux entering the sheath at the sheath boundary is then

\begin{equation}
    \Gamma_{e,\mathrm{sb}}
    =
    n_{e,\mathrm{sb}}
    \left\langle v_x^+ \right\rangle ,
    \label{eq:incoming_electron_flux}
\end{equation}

where $n_{e,\mathrm{sb}}$ is the electron density obtained from the fluid solution at the sheath boundary, $ n_{e,\mathrm{sb}} = n_{e,\mathrm{ref}}\exp(\Psi_{\mathrm{sb}})$. The ion flux entering the sheath and reaching the wall is given by
\begin{equation}
    \Gamma_i =
    n_{i,sb} \mathcal{U}_{\mathrm{sb}}v_B =
    n_{e,\mathrm{ref}}v_B,
    \label{eq:ion_flux_coupling}
\end{equation}
with $n_{i,sb}$ the ion density at the sheath boundary. 
Equations \ref{eq:incoming_electron_flux} and \ref{eq:ion_flux_coupling} define the electron and ion fluxes used in the wall potential search.

To determine the normalized wall potential while accounting for the effects of electron transport through the sheath and the non-Maxwellian EEDF, the fluid equations described above are integrated from the initial conditions through the sheath until sufficiently negative potentials are reached, i.e., $\Psi(x)=-10$. The sheath boundary is identified by Eq.~\ref{eq:u_s_velocity}, and the wall is placed at the position at which the normalized potential,  $\Phi(x)=e\phi(x)/k_B T_e=\Psi(x)-\Psi_{sb}$, referenced to the sheath boundary, reaches an initial estimate for the classical floating-potential estimate based on a Maxwellian EEDF:

\begin{equation}
    \Phi_w^{(0)}
    =
    \frac{1}{2}\ln\left(\frac{2\pi m_e}{M_i}\right).
    \label{eq:phi_wall_Maxwellian}
\end{equation} 

The potential and electric field between these two positions are interpolated onto the 25-cell grid used by the electron kinetic simulation. A pilot electron simulation is then performed in this trial field, following electrons until 1000 electrons reach the wall. From the pilot simulations, we determine the electron transmission through the sheath: If $N_w$ is the number of electrons that reach the wall and $N_{\mathrm{inj}}$ the number of injected electrons, then the electron flux at the wall is:

\begin{equation}
    \Gamma_{e,w}
    =
    \Gamma_{e,\mathrm{sb}}
    \frac{N_w}{N_{\mathrm{inj}}}.
    \label{eq:electron_wall_flux}
\end{equation}

To achieve flux balance at the wall, $\Gamma_{e,w}=\Gamma_i$, the relative flux error, $\delta_\Gamma = \frac{\Gamma_{e,w}-\Gamma_i}{\Gamma_i}$, needs to become zero. After each pilot simulation, $\delta_\Gamma$ is calculated and the wall potential, and with it the wall position, is updated according to

\begin{equation}
    \Phi_w^{(k+1)}
    =
    \Phi_w^{(k)}-0.4\,\delta_\Gamma.
    \label{eq:wall_potential_update}
\end{equation}

If too many electrons reach the wall, Eq.~\ref{eq:wall_potential_update} makes the wall potential more negative and reduces the electron flux, and vice versa. The factor 0.4 was chosen empirically. Iteration of the wall potential is continued until $\left| \delta_\Gamma \right| < 0.04$, evaluated at 1000 wall hits.

Once this criterion is satisfied, the wall potential, sheath thickness, and electric-field profile are fixed. A full electron simulation is initiated using the same flux-weighted injection distribution, Equation~\ref{eq:flux_weighted_injection}. In the present calculations, electrons were injected until 10,000 electrons reached the wall. 

Various tests performed to validate the numerical code are described in the Supporting Information, section S1 and Figure S1. 

\section{Results and Discussion}

In discussions of electron kinetics, it is common to focus on the electron energy probability function (EEPF), $F(u)$, which is related to the EEDF by $\mathcal{F}(u) = F(u) \sqrt{u}$. When plotted as a function of the kinetic energy $u$, we use the common normalization $\int_0^\infty F(u) \sqrt{u}du = 1$.

A series of simulations was performed for pure helium and helium--water vapor mixtures containing 96.48\% He and 3.52\% H$_2$O, assuming a gas temperature of 300 K, a total pressure of 101.3 kPa, and the mixture ratio being based on the saturation vapor pressure of water at 300 K (27 $^\circ$C). The reduced electric field strength, $E/N$, was varied from 10 to 40~Td, while the ion density at the sheath boundary, $n_{i,sb}$, was varied from $10^{17}$ to $10^{21}$~m$^{-3}$. In the following discussion, we focus on two cases: pure helium at $E/N = 15$~Td and the He/H$_2$O mixture at $E/N = 40$~Td. These conditions were selected because they yield bulk plasma EEPFs with comparable mean electron kinetic energies. Results for the remaining simulation conditions are presented in the Supporting Information, Table S1 and Figures S2-S11.

\begin{table}[htbp]
  \centering
  \caption{Sheath thickness, normalized wall potential, and wall ion velocity normalized to the Bohm velocity for pure He at $E/N=15\,\mathrm{Td}$ and He/H$_2$O (96.48\%/3.52\%) at $E/N=40\,\mathrm{Td}$ as functions of ion density at the sheath boundary $n_{i,sb}$. Here, $|\Phi_w|=e|\phi_w|/k_BT_e$ is the wall potential normalized to $k_B T_e/e$, and $v_{i,w}/v_\mathrm{B}$ is the normalized wall ion velocity.}
  \label{tab:sheath-summary-15-40td}
  \begin{tabular}{c S[table-format=3.2] S[table-format=1.2] S[table-format=1.3] S[table-format=3.2] S[table-format=1.2] S[table-format=1.3]}
    \toprule
    & \multicolumn{3}{c}{Pure He} & \multicolumn{3}{c}{He/H$_2$O (96.48\%/3.52\%)} \\
    \cmidrule(lr){2-4}\cmidrule(lr){5-7}
    $n_{i,sb}$ (m$^{-3}$) & {$L_s$ ($\mu$m)} & {$|\Phi_w|$} & {$v_{i,w}/v_\mathrm{B}$} & {$L_s$ ($\mu$m)} & {$|\Phi_w|$} & {$v_{i,w}/v_\mathrm{B}$} \\
    \midrule
    $10^{17}$ & 167 & 4.33 & 0.0301 & 147 & 3.59 & 0.0279 \\
    $10^{18}$ & 68 & 4.31 & 0.0798 & 68.6 & 4.37 & 0.0807 \\
    $10^{19}$ & 26.6 & 3.96 & 0.203 & 28.3 & 4.43 & 0.216 \\
    $10^{20}$ & 10.5 & 3.65 & 0.5 & 10.9 & 3.97 & 0.529 \\
    $10^{21}$ & 4.08 & 3.47 & 1.1 & 4.12 & 3.55 & 1.12 \\
    \bottomrule
  \end{tabular}
\end{table}

Table~\ref{tab:sheath-summary-15-40td} summarizes the results of the sheath simulations for pure helium and the He/H$_2$O mixture. For pure helium, the sheath thickness decreases from 167~$\mu$m at $n_{i,sb} = 10^{17}$~m$^{-3}$ to approximately 4.1~$\mu$m at $n_{i,sb} = 10^{21}$~m$^{-3}$. The magnitude of the normalized sheath potential decreases from 4.33 to 3.47. While these values are close to the classical value of 3.52 for a Maxwellian electron distribution, the deviations indicate the influence of the non-Maxwellian character of the EEPF. Remarkably, the ion velocity at the wall is significantly lower than the Bohm velocity for the four lowest densities. Only at $n_{i,sb} = 10^{21}$~m$^{-3}$ does the ion velocity at the wall slightly exceed the Bohm velocity. This behavior is characteristic of a highly collisional sheath, in which ions do not reach the Bohm velocity at the sheath boundary.

For the He/H$_2$O mixture, the sheath thicknesses are similar to those obtained for pure helium. The magnitude of the normalized wall potential increases from 3.59 at $n_{i,sb} = 10^{17}$~m$^{-3}$ to a maximum of 4.43 at $n_{i,sb} = 10^{19}$~m$^{-3}$, before decreasing to 3.55 at $n_{i,sb} = 10^{21}$~m$^{-3}$. The origin of this maximum remains to be explored further. However, similar behavior is observed in additional simulations for He/H$_2$O mixtures presented in the Supporting Information, suggesting that the maximum is not an artifact of statistical noise in the simulations. As in pure helium, the ion velocity at the wall remains significantly below the Bohm velocity for all but the highest plasma density.

\begin{figure}
    \centering
    \includegraphics[width=1\linewidth]{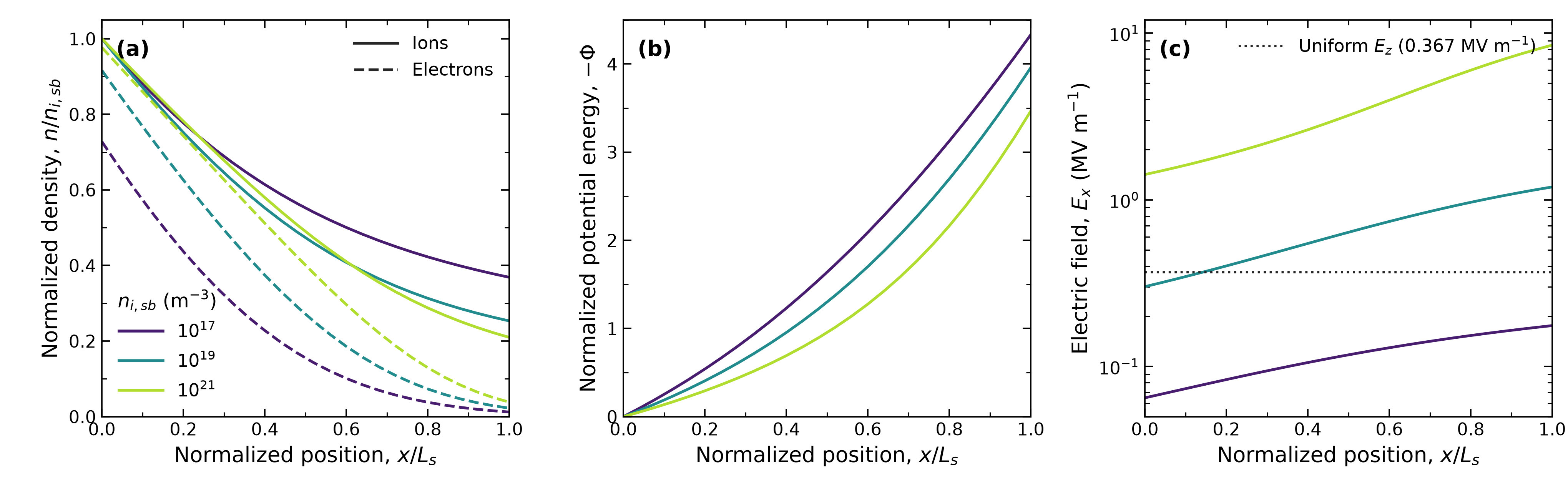}
    \caption{Plasma properties in the sheath. (a) Ion and electron densities from the fluid sheath model, normalized to the ion density at the sheath boundary. (b) Electron potential energy $e|\phi|$ normalized to $k_B T_e$ of the bulk EEPF. (c) Sheath electric field, $E_x$, compared to the electric field strength $E_z$ in the bulk plasma.}
    \label{fig:sheath_profiles_He_15Td}
\end{figure}

Figure~\ref{fig:sheath_profiles_He_15Td} shows various sheath-model properties for pure helium. Figure~\ref{fig:sheath_profiles_He_15Td}(a) depicts the electron and ion density profiles. At the lowest sheath boundary density of $n_{i,sb} = 10^{17}$~m$^{-3}$, a significant deviation from quasineutrality is observed at the sheath boundary, according to the definition by Godyak and Sternberg. This deviation is smaller than that obtained using the original sheath boundary definition in Kono's model and is characteristic of a highly collisional sheath. As sheath-boundary density increases and the sheath becomes thinner and less collisional, the deviation from quasineutrality at the sheath boundary decreases, and at $n_{i,sb} = 10^{21}$~m$^{-3}$, quasineutrality is nearly achieved.

Figure~\ref{fig:sheath_profiles_He_15Td}(b) shows the electron potential energy within the sheath, normalized to $k_BT_e$ of the bulk EEDF. At $n_{i,sb} = 10^{17}$~m$^{-3}$, the wall potential energy is significantly higher than the value predicted by the classical expression assuming a Maxwellian electron distribution, Equation~\ref{eq:phi_wall_Maxwellian}. As the sheath boundary density increases, the electron potential energy at the wall decreases.

Figure~\ref{fig:sheath_profiles_He_15Td}(c) shows the sheath electric field strength, $E_x$. At the lowest density, the sheath electric field is weaker than the applied electric field $E_z$ because the sheath is relatively thick. At $n_{i,sb} = 10^{19}$~m$^{-3}$, the sheath electric field is of the same order of magnitude as, or somewhat stronger than, the applied electric field. At the highest sheath boundary density, the reduced sheath thickness yields significantly stronger sheath electric fields, with $E_x$ up to about 20 times $ E_z$.

\begin{figure}
    \centering
    \includegraphics[width=1\linewidth]{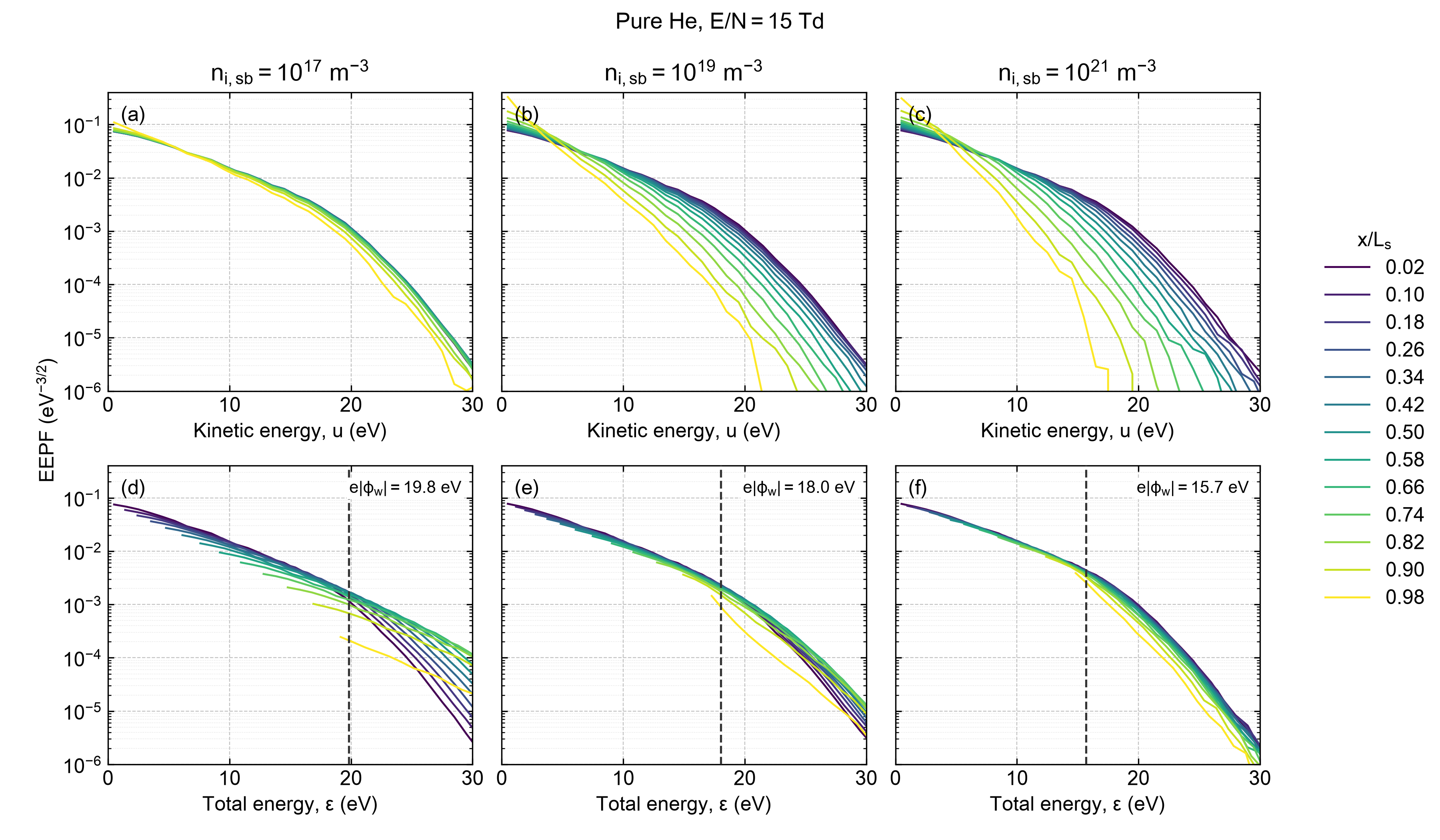}
    \caption{Electron energy probability functions (EEPFs) for pure helium at $E/N = 15$~Td for sheath boundary densities of $n_{i,sb} = 10^{17}$, $10^{19}$, and $10^{21}$~m$^{-3}$ (left to right). The EEPFs are shown at different positions within the sheath as functions of (a)--(c) electron kinetic energy and (d)--(f) total electron energy. At low plasma density, the EEPFs exhibit local behavior, remaining nearly independent of position when plotted against kinetic energy. At higher densities, the EEPFs approach the nonlocal limit, collapsing onto a nearly universal distribution when plotted against total energy, except near the absorbing wall.}
    \label{fig:EEPF_He_15Td}
\end{figure}

Before discussing the EEDFs in the sheath, we review two common approximations of solving the Boltzmann equation in low temperature plasmas:
\begin{itemize}
    \item The local field approximation (LFA): As the name implies, the LFA assumes that all plasma spatial nonuniformities can be neglected, and that the EEDF is only determined by the local electric field strength at the position considered. Common Boltzmann solvers such as BOLSIG+ \cite{hagelaarSolvingBoltzmannEquation2005} and LoKI-B \cite{tejero-del-cazLisbOnKIneticsBoltzmann2019} rely on the LFA.
    \item The Nonlocal Approximation (NLA): Under conditions when spatial electron diffusion is much faster than electron energy change, either through energy gain by heating or energy loss by collisions, the EEDF is appropriately described by a single distribution function of total electron energy, $\varepsilon = u + e|\phi(x)|$ \cite{tsendinEnergyDistributionElectrons1974,tsendinElectronKineticsNonuniform1995a,kolobovAdvancesElectronKinetics2013}. This reflects that electrons travel through the system, here the sheath, with essentially constant $\varepsilon$. Because of the requirement of fast electron diffusion, the applicability of the NLA is most commonly associated with low pressures \cite{godyakParadoxicalSpatialDistribution1993,kortshagenElectronIonDistribution1995,kaganovichSpacetimeaveragingProcedureModeling1992,berezhnoiFastModellingLowpressure1998,loffhagen_electron_2005,singhMeasurementsElectronEnergy2000a,loffhagen_multiterm_2022,kimTwodimensionalAnalysisTransition2022,wenNonlocalElectronDynamics2025, czarnetzkiDescribingLocalNonlocal2022}. However, nonlocal effects were also observed in the electron spectrum of microdischarges in He and Ar at pressures as high as about 20 Torr \cite{kudryavtsev_registration_2012,kudryavtsev_use_2015}. 
\end{itemize}

Figure~\ref{fig:EEPF_He_15Td} shows the EEPFs for pure helium at $E/N = 15$~Td for three different sheath boundary densities. The upper row shows the EEPFs, normalized to unity, at various sheath positions as functions of electron kinetic energy, $u$. The lower row, Figure~\ref{fig:EEPF_He_15Td}(d)--(f), shows the same EEPFs, normalized to the electron density at the sheath boundary, plotted as functions of total electron energy $\varepsilon$. 

At the lowest sheath boundary density, $n_{i,sb} = 10^{17}$~m$^{-3}$, Figure~\ref{fig:EEPF_He_15Td}(a), i.e., the thickest sheath, the EEPF vs. kinetic energy remains essentially unchanged throughout the sheath, consistent with the LFA and $E_z = const \gg E_x$. At the intermediate sheath boundary density of $n_{i,sb} = 10^{19}$~m$^{-3}$, shown in Figure~\ref{fig:EEPF_He_15Td}(b), significant deviations from the LFA are observed, with the EEPFs near the wall exhibiting lower mean electron energies than those near the sheath boundary. This effect becomes even more pronounced at the highest sheath-boundary density, $n_{i,sb} = 10^{21}$~m$^{-3}$, i.e., the thinnest sheath, shown in Figure~3(c). When plotted against total energy $\varepsilon$, the EEPFs show the opposite trend. At the lowest sheath boundary density, Figure~\ref{fig:EEPF_He_15Td}(d), the EEPFs plotted against total energy vary significantly throughout the sheath; at the intermediate and highest densities, Figure~\ref{fig:EEPF_He_15Td}(e) and (f), the EEPFs collapse almost entirely onto a single universal curve, the hallmark of the NLA. The only notable exceptions occur in the cells closest to the wall, where electron losses at the absorbing surface deplete the EEPF, as was also reported in a previous numerical study of nonlocal electron kinetics \cite{kortshagenComparisonMonteCarlo1996}. 

\begin{figure}
    \centering
    \includegraphics[width=1\linewidth]{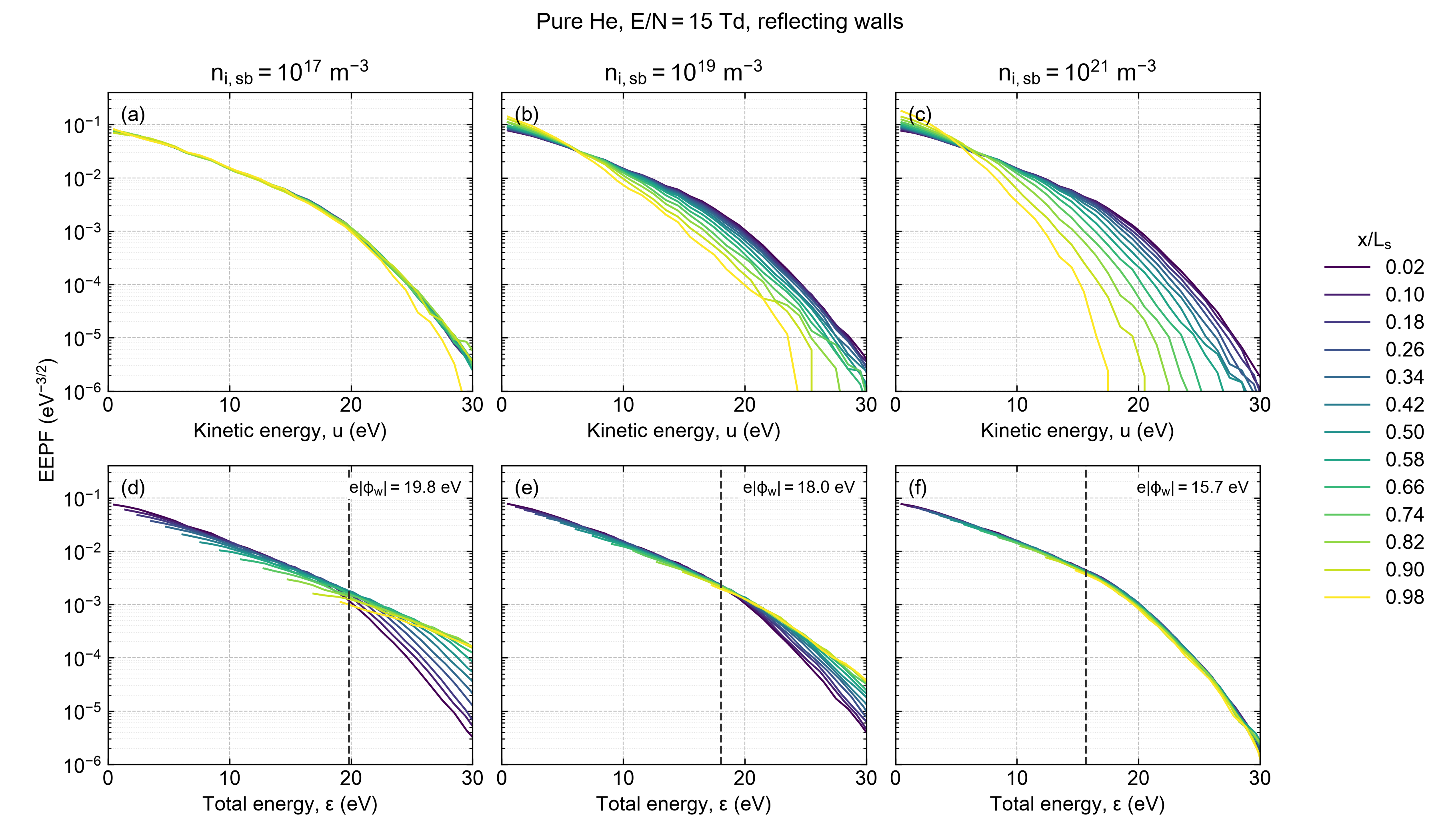}
    \caption{Electron energy probability functions (EEPFs) for pure helium at $E/N = 15$~Td for the same conditions as in Figure~\ref{fig:EEPF_He_15Td}, but with fully reflecting walls.}
    \label{fig:EEPF_He_15Td_reflecting_walls}
\end{figure}

The effect of wall losses is illustrated by Figure~\ref{fig:EEPF_He_15Td_reflecting_walls}, where simulations were run for a fully reflecting wall and otherwise the same conditions as in Figure~\ref{fig:EEPF_He_15Td}, including the same sheath potentials. The removal of the electron sink at the wall leads to a significant increase of the EEPFs in the cells close to the wall. In the case of the thinnest sheath, Figure~\ref{fig:EEPF_He_15Td_reflecting_walls}(f), the EEPFs plotted against total energy are now fully collapsed into a single curve, consistent with the classic NLA. However, the removal of wall losses leads to changes of the EEPFs for all sheath boundary densities. Even for the thickest sheath, Figures ~\ref{fig:EEPF_He_15Td_reflecting_walls}(a) and ~\ref{fig:EEPF_He_15Td_reflecting_walls}(d), the EEPF is significantly changed for the last 10 \% of the sheath in front of the wall.    

Overall, the finding that nonlocal EEPFs occur at pressures as high as atmospheric pressure may appear counterintuitive, but it is consistent with the very thin sheaths observed at the higher sheath boundary densities. We discuss the relationship between sheath thickness, $L_s$, and energy relaxation length, $\lambda_\varepsilon$, below.

\begin{figure}
    \centering
    \includegraphics[width=1\linewidth]{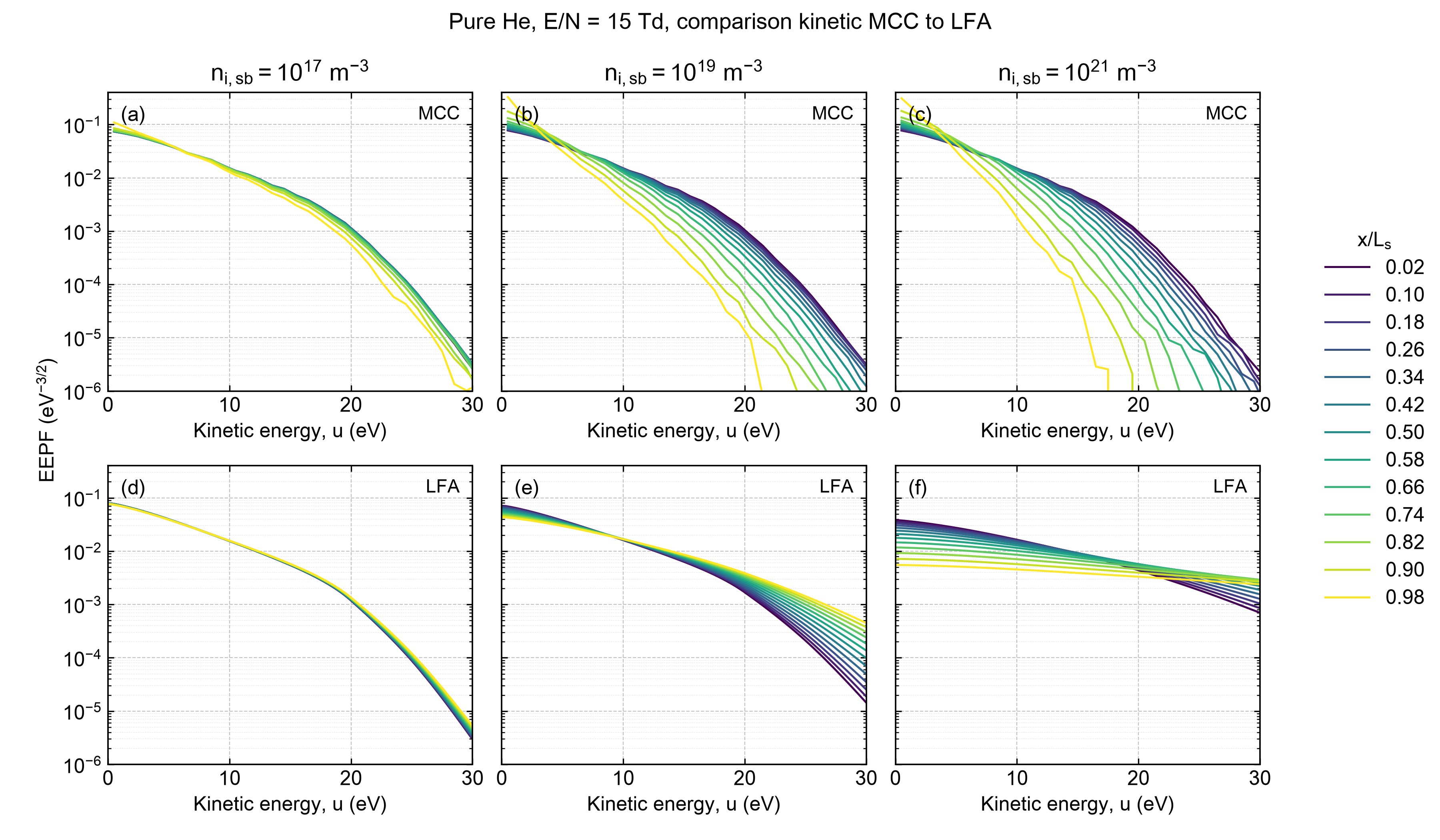}
    \caption{Comparison of electron energy probability functions (EEPFs) for pure helium at $E/N = 15$~Td for sheath boundary densities of $n_{i,sb} = 10^{17}$, $10^{19}$, and $10^{21}$~m$^{-3}$ (left to right) obtained from the particle kinetic Monte Carlo collision (MCC) simulation (a)-(c), and the local field approximation (LFA) which includes the sheath electric field as an electron heating component (d)-(f). In particular, at sheath boundary densities of $10^{19}$ and $10^{21}$~m$^{-3}$, the LFA including the sheath field shows unreasonably large electron energies, in contradiction to the kinetic simulation.}
    \label{fig:EEPF_LFA}
\end{figure}

The applicability of the LFA requires further discussion in view of the sheath and applied electric fields shown in Figure~\ref{fig:sheath_profiles_He_15Td}(c). The validity of the LFA in Figure~\ref{fig:EEPF_He_15Td}(a) is consistent with the applied electric field $E_z$ being much stronger than the sheath electric field $E_x$. However, for the conditions shown in Figures~\ref{fig:EEPF_He_15Td}(b) and (c), the sheath electric field $E_x$ exceeds the applied electric field $E_z$. Nevertheless, the EEPFs exhibit a decrease in electron energy as the wall is approached. This behavior correctly reflects the physical nature of the sheath electric field $E_x$, which acts to decelerate electrons approaching the wall. A physically incorrect use of the LFA, in which the sheath electric field is incorrectly used as part of the heating electric field strength, $E_{\mathrm{heating}} = \sqrt{E_z^2 + E_x^2}$, would predict increasingly energetic EEPFs toward the wall, in direct contradiction to the kinetic simulation results, as demonstrated in Figure~\ref{fig:EEPF_LFA}.

\begin{figure}
    \centering
    \includegraphics[width=1\linewidth]{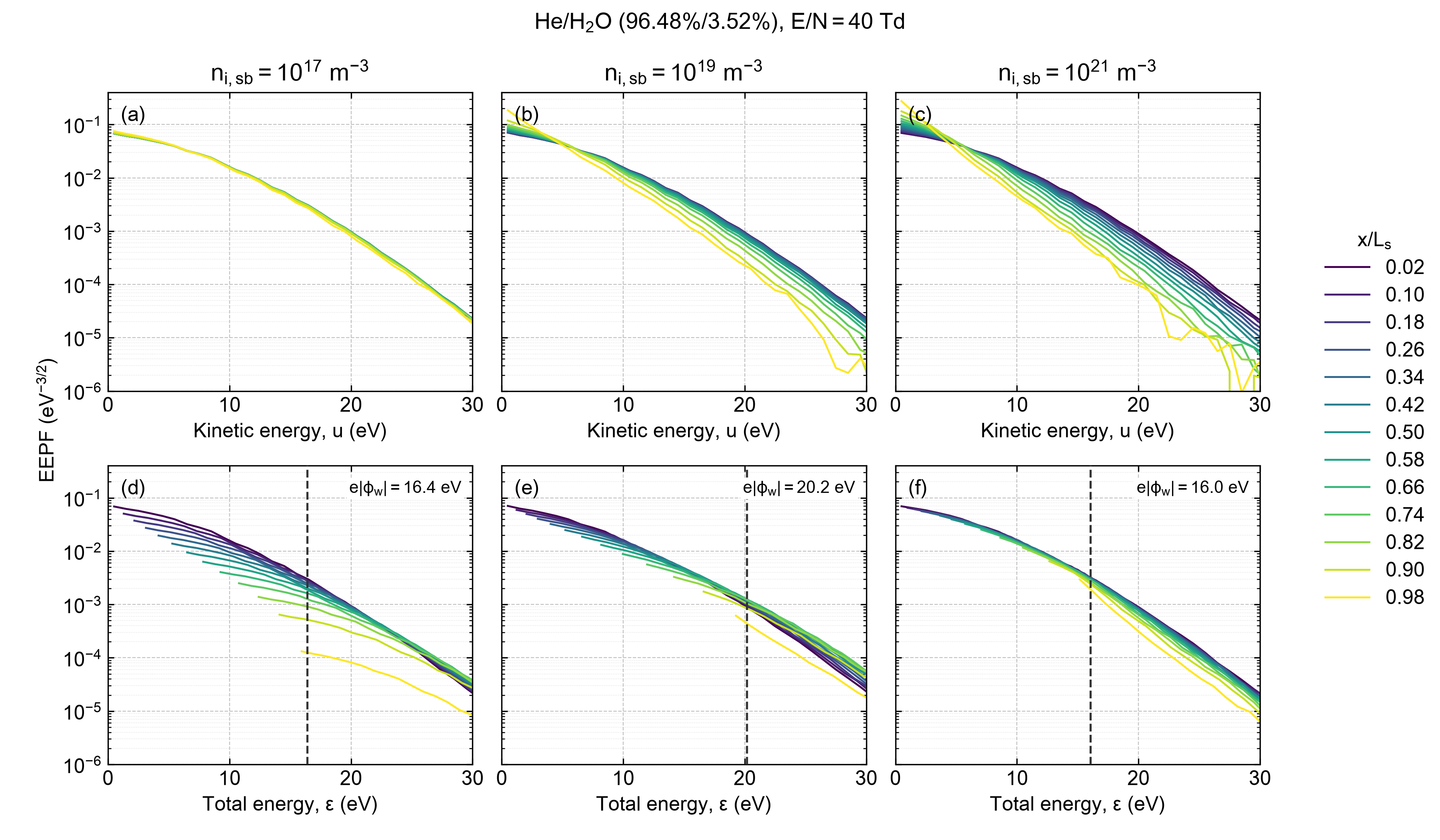}
    \caption{Electron energy probability functions (EEPFs) for a He/H$_2$O mixture (96.48\%/3.52\%) at $E/N = 40$~Td for sheath boundary densities of $n_{i,sb} = 10^{17}$, $10^{19}$, and $10^{21}$~m$^{-3}$ (left to right). The EEPFs are shown at different positions within the sheath as functions of (a)--(c) electron kinetic energy and (d)--(f) total electron energy. At low plasma density, the EEPFs exhibit local behavior, remaining nearly independent of position when plotted against kinetic energy. At higher densities, the EEPFs approach the nonlocal limit, collapsing onto a nearly universal distribution when plotted against total energy, except near the absorbing wall.}
    \label{fig:EEPF_He_H2O_40Td}
\end{figure}

Figure~\ref{fig:EEPF_He_H2O_40Td} shows the EEPFs for the He/H$_2$O mixture. At the lowest sheath boundary density of $n_{i,sb} = 10^{17}$~m$^{-3}$, the EEPFs remain consistent with the LFA. At the highest sheath boundary density of $n_{i,sb} = 10^{21}$~m$^{-3}$, the EEPFs again approach the nonlocal limit. However, at the intermediate density of $n_{i,sb} = 10^{19}$~m$^{-3}$, the EEPFs remain closer to the LFA than in the corresponding pure helium case. This indicates that the additional electron collision processes introduced by water vapor, particularly vibrational excitation, shift the electron kinetics toward the local field regime, as discussed further below.

\begin{figure}
    \centering
    \includegraphics[width=0.5\linewidth]{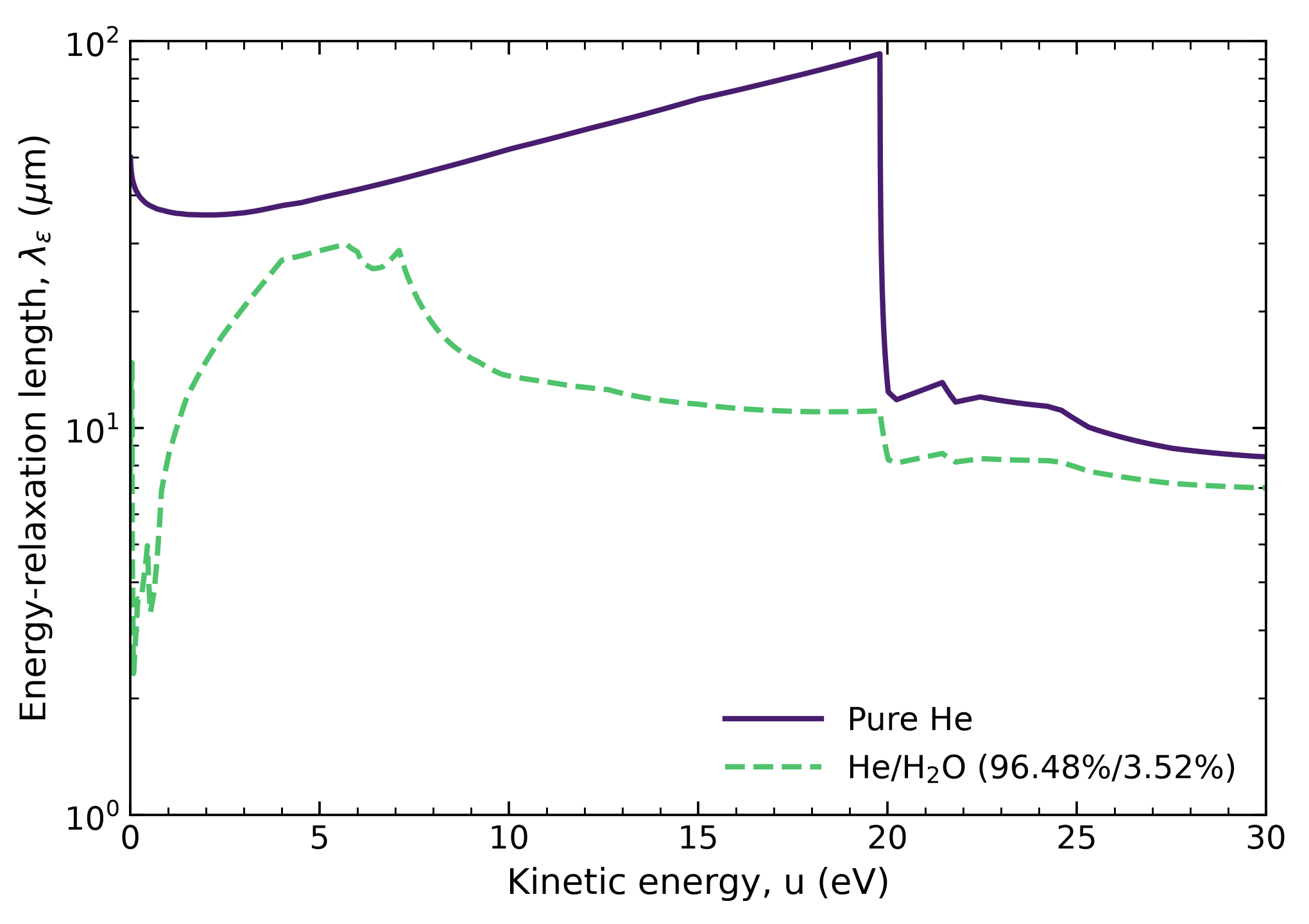}
    \caption{Electron energy relaxation length, $\lambda_\varepsilon$, as a function of electron kinetic energy for pure He and a He/H$_2$O mixture (96.48\%/3.52\%) at atmospheric pressure. The relaxation lengths were calculated using Eq.~\ref{eq:lambda_epsilon}, accounting for elastic and inelastic electron-neutral collisions. The addition of water vapor significantly reduces $\lambda_\varepsilon$ at energies below the first excitation threshold of He due to additional inelastic energy loss processes, shifting the electron kinetics toward the local field regime.}
    \label{fig:lambda_epsilon}
\end{figure}

The transition behavior observed in Figures~\ref{fig:EEPF_He_15Td} and ~\ref{fig:EEPF_He_H2O_40Td} can be understood by considering the energy relaxation length, $\lambda_{\varepsilon}$ \cite{tsendinEnergyDistributionElectrons1974,tsendinElectronKineticsNonuniform1995a,kolobovAdvancesElectronKinetics2013}, which describes the length over which electron energy changes significantly due to collisions. $\lambda_{\varepsilon}$ for He and the He/H$_2$O mixture (96.48\%:3.52\%) is shown in Figure ~\ref{fig:lambda_epsilon}. It was calculated from

\begin{equation}
\lambda_{\varepsilon} = \lambda_m \left[ \frac{\sum_s \chi_s \nu_{m,s}^{(0)}}{\displaystyle \sum_s \chi_s \left( \frac{2m_e}{M_s}\nu_{m,s}^{(0)} + \sum_{j\in s} \frac{u_{\mathrm{loss},j}}{u} \nu_{j,s}^{(0)} \right)} \right]^{1/2},
\label{eq:lambda_epsilon}
\end{equation}

where $\chi_s$ and $M_s$ are the molar fraction and molecular mass of species $s$, respectively, $\nu_{m,s}^{(0)}$ is its momentum-transfer collision frequency evaluated using the total neutral density, and $\nu_{j,s}^{(0)}$ is the corresponding collision frequency for inelastic process $j$. The momentum-relaxation mean free path is $\lambda_m=[N\sum_s\chi_s\sigma_{m,s}]^{-1}$, with $N$ the total gas density and $\sigma_{m,s}$ the momentum transfer cross section of species $s$. The energy loss $u_{\mathrm{loss},j}$ was taken as the threshold energy for excitation, dissociation, and ionization. Attachment was treated as the loss of the electron's complete energy.

For pure He, the electron energy relaxation length $\lambda_\varepsilon$ is on the order of several tens of micrometers at energies below the first excitation threshold. At energies where inelastic collisions become possible, $\lambda_\varepsilon$ decreases to approximately 10~$\mu$m. Comparing these values with the sheath thicknesses listed in Table~1 explains the observed transition between local and nonlocal electron kinetics. At the lowest sheath boundary density, the sheath thickness significantly exceeds the energy relaxation length, consistent with the LFA. At the highest sheath boundary density, the sheath thickness is considerably smaller than $\lambda_\varepsilon$, explaining the observed nonlocal behavior. At the intermediate density of $n_{i,sb} = 10^{19}$~m$^{-3}$, the sheath thickness is comparable to $\lambda_\varepsilon$, accounting for the intermediate behavior observed in the EEPFs.

The addition of water vapor reduces $\lambda_\varepsilon$ primarily in the energy range where electrons undergo only elastic collisions in pure He. This reduction results from additional inelastic collision processes with threshold energies below the first excitation threshold of He, including vibrational excitation, dissociative attachment, dissociation, and ionization. The shorter energy relaxation length explains the more localized behavior of the EEPFs observed for the He/H$_2$O mixture in Figure~\ref{fig:EEPF_He_H2O_40Td}.

\begin{figure}
    \centering
    \includegraphics[width=1\linewidth]{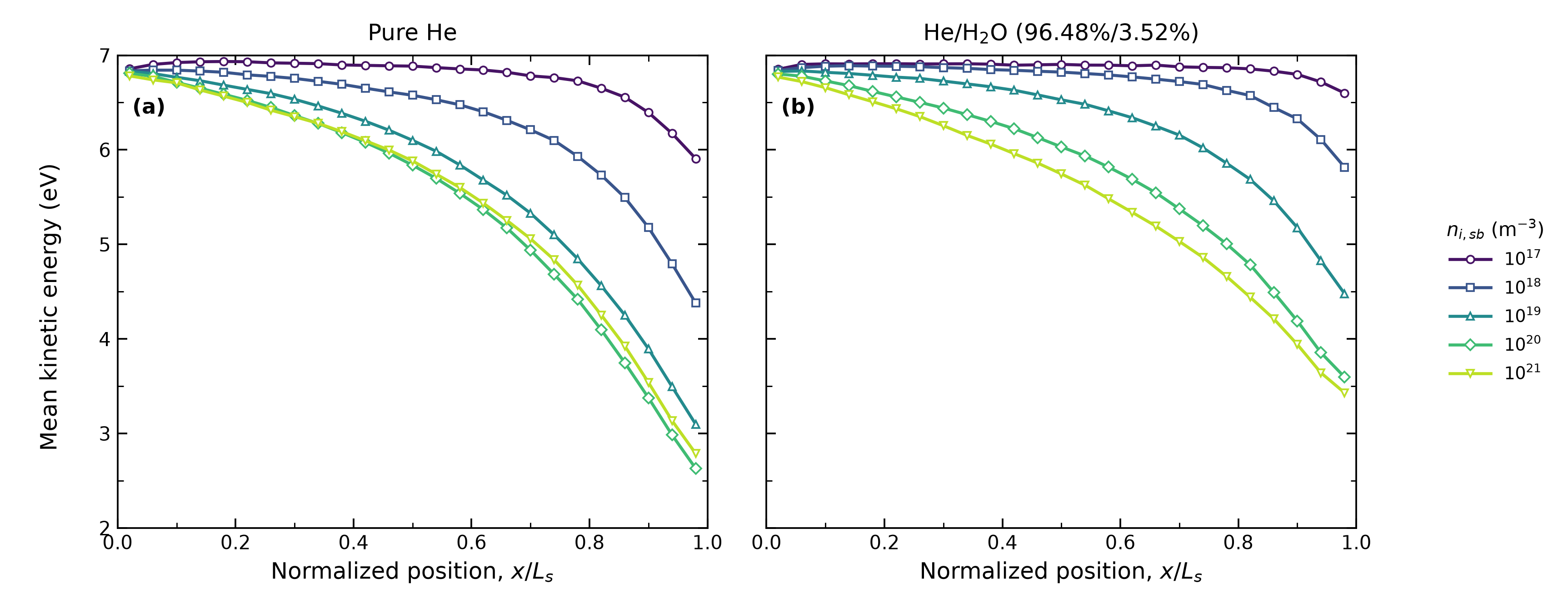}
    \caption{Spatial variation of the electron mean kinetic energy within the sheath for (a) pure He at $E/N = 15$~Td and (b) a He/H$_2$O mixture (96.48\%/3.52\%) at $E/N = 40$~Td for different sheath boundary densities $n_{i,sb}$. The position is normalized to the sheath thickness $L_s$, with $x/L_s = 0$ corresponding to the sheath boundary and $x/L_s = 1$ to the wall. At low plasma densities, the mean kinetic energy remains nearly constant throughout the sheath, whereas at higher densities it decreases significantly toward the wall. The addition of water vapor reduces the spatial variation of the mean kinetic energy, consistent with the shorter electron energy relaxation length and more localized electron kinetics.}
    \label{fig:u_mean}
\end{figure}

Figure~\ref{fig:u_mean} shows the spatial variation of the electron mean kinetic energy within the sheath for pure He and the He/H$_2$O mixture. For pure He, shown in Figure~\ref{fig:u_mean}(a), the mean kinetic energy remains relatively constant throughout most of the sheath at the lowest sheath boundary density, decreasing only in the vicinity of the wall. With increasing sheath boundary density, the spatial variation of the mean kinetic energy becomes more pronounced. At the highest sheath boundary density of $n_{i,sb} = 10^{21}$~m$^{-3}$, the mean kinetic energy decreases from approximately 6.8~eV in the plasma to 2.8~eV at the wall. The addition of water vapor, shown in Figure~\ref{fig:u_mean}(b), results in a more uniform mean kinetic energy throughout the sheath. At the lowest sheath boundary density, the mean kinetic energy remains essentially constant, with only a slight decrease within the last 10\% of the sheath thickness near the wall. At the highest sheath boundary density, the mean kinetic energy decreases from approximately 6.8~eV in the plasma to 3.5~eV at the wall. The less pronounced spatial variation of the mean kinetic energy in the presence of water vapor is consistent with the more localized behavior of the EEPF resulting from the reduced electron energy relaxation length.

\begin{figure}
    \centering
    \includegraphics[width=1\linewidth]{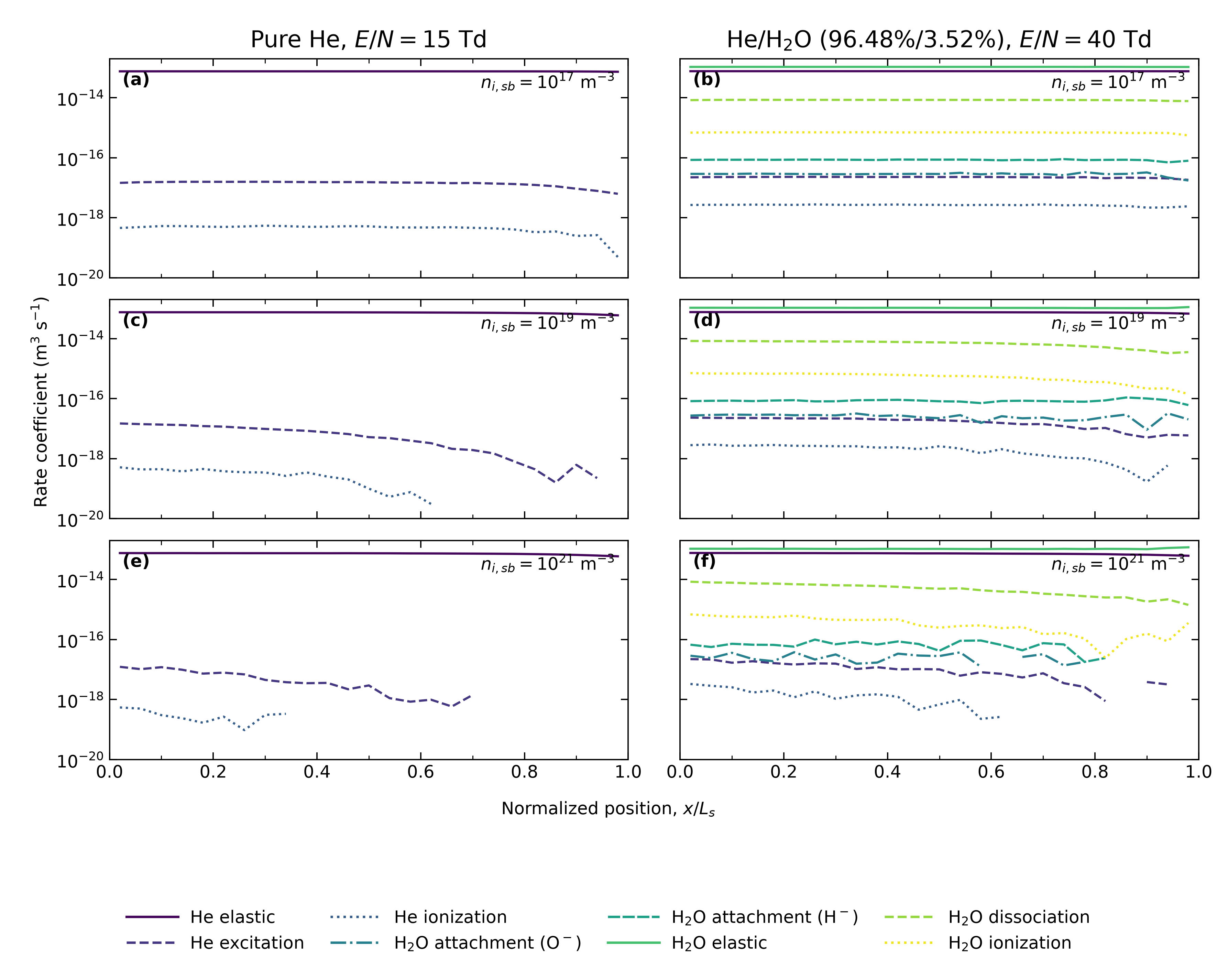}
\caption{Spatial variation of electron-impact reaction rate coefficients within the sheath for (a), (c), (e) pure He at $E/N = 15$~Td and (b), (d), (f) a He/H$_2$O mixture (96.48\%/3.52\%) at $E/N = 40$~Td for sheath boundary densities of $n_{i,sb} = 10^{17}$, $10^{19}$, and $10^{21}$~m$^{-3}$ (top to bottom). The position is normalized to the sheath thickness $L_s$, with $x/L_s = 0$ corresponding to the sheath boundary and $x/L_s = 1$ to the wall. At low plasma densities, the rate coefficients remain nearly constant throughout the sheath, whereas at higher densities, the rates of high-threshold inelastic processes decrease significantly toward the wall. The addition of water vapor reduces these spatial variations, consistent with the shorter electron energy relaxation length and more localized electron kinetics. Missing portions of individual curves indicate spatial regions where no corresponding collision events were recorded in the particle kinetic Monte Carlo simulations.}
\label{fig:rate_coefficients}
\end{figure}

Figure~\ref{fig:rate_coefficients} shows the spatial variation of the electron-impact reaction rate coefficients derived from the EEPFs presented in Figures~\ref{fig:EEPF_He_15Td} and \ref{fig:EEPF_He_H2O_40Td}. For pure He, the rate coefficients remain relatively constant throughout the sheath at the lowest sheath boundary density, with only a slight decrease in the rates of inelastic processes near the wall. At the highest sheath boundary density, the rate coefficients for excitation and ionization decrease significantly toward the wall. These coefficients cannot be plotted all the way to the wall because no excitation or ionization events were observed in the corresponding spatial cells. Before reaching this cutoff, the rate coefficients decrease by approximately an order of magnitude. Similar behavior is observed at the intermediate sheath boundary density, where the excitation rate coefficient decreases by almost two orders of magnitude from the sheath boundary toward the wall.

For the He/H$_2$O mixture, the rate coefficients at the lowest sheath boundary density remain nearly constant throughout the sheath, consistent with the local field behavior of the EEPF. This is also observed at the intermediate sheath boundary density, consistent with the more localized electron kinetics resulting from the reduced electron energy relaxation length. At the highest sheath boundary density, the rate coefficients for processes with the highest threshold energies exhibit the strongest spatial variations, with the rate coefficients for He excitation and ionization decreasing by more than an order of magnitude from the sheath boundary toward the wall.

Overall, the behavior of the rate coefficients demonstrates that the transition from local to nonlocal electron kinetics leads to significant modifications of the electron-impact reaction kinetics within the sheath, even at pressures as high as atmospheric pressure.

\section{Conclusions}

We have investigated the electron kinetics within the sheath and at the plasma-surface interface of atmospheric pressure plasmas in pure helium and helium-water vapor mixtures. For this purpose, we developed a hybrid model that combines a fluid description of the collisional plasma sheath with a kinetic Monte Carlo model for electrons. The model accounts for the non-Maxwellian electron energy distribution in the bulk plasma, electron collisions and acceleration within the sheath, and electron losses to the wall. 

Our results demonstrate that electron kinetics within the sheath can exhibit both local and nonlocal behavior, even at atmospheric pressure. The transition between these regimes is determined primarily by the ratio of the electron energy relaxation length, $\lambda_\varepsilon$, to the sheath thickness, $L_s$. At low plasma densities, where $L_s \gg \lambda_\varepsilon$, the electron energy probability function (EEPF) remains nearly independent of position when expressed as a function of kinetic energy, consistent with the local field approximation. At high plasma densities, where $L_s \ll \lambda_\varepsilon$, the EEPFs approach a universal distribution when expressed as functions of total electron energy, characteristic of nonlocal electron kinetics. The addition of water vapor reduces the energy relaxation length through additional inelastic electron collisions and shifts the electron kinetics toward the local regime.

In the nonlocal regime, the electron mean kinetic energy decreases significantly toward the wall. For the highest sheath boundary density investigated, $n_{i,sb} = 10^{21}$~m$^{-3}$, the mean electron kinetic energy decreases from approximately 6.8~eV to 2.8~eV in pure helium and to 3.5~eV in the helium-water vapor mixture. This decrease is accompanied by significant changes in electron-impact reaction rate coefficients, particularly for processes with high threshold energies. Excitation and ionization rate coefficients can decrease by more than an order of magnitude toward the wall, demonstrating that the electron-impact reaction kinetics within the sheath can differ substantially from those in the bulk plasma.

These findings have important implications for the modeling of low-temperature plasmas interacting with solid and liquid surfaces. In particular, they demonstrate that the high collision frequencies at atmospheric pressure do not necessarily justify the application of the local field approximation within the sheath. Moreover, the sheath electric field cannot simply be included as an additional electron-heating field in conventional Boltzmann solvers, since its primary role is to decelerate electrons approaching the wall. Accurate modeling of electron kinetics near plasma-surface interfaces therefore requires accounting for the sheath electric field, nonlocal electron transport, and electron losses to the surface. This is particularly important for applications in which electron-induced reactions at or near the plasma-surface interface play a central role.

\ack{The author acknowledges Professors Marien Simeni Simeni and Suo Yang for critically reading the manuscript and providing feedback. 

During the preparation of this work, the author used
ChatGPT and Codex (OpenAI) for code development support, including review, testing, and code implementation, data analysis, and figure preparation. ChatGPT and Grammarly were used in writing, reviewing, and editing the manuscript. All outputs were critically
reviewed and modified as necessary by the author. The author takes full responsibility for the content of the publication.}

\funding{Initial work was sponsored by the Army Research Office  under Grant No. W911NF-20-1-0105. The work was continued with support from the U.S. Department of Energy, Office of Science, under grant DE-SC0025639.}

\roles{Uwe Kortshagen performed the work described in this article, including conceptualization, code development, data analysis and curation, and writing of the manuscript.}

\data{All data shown in the main text and supporting information will be made available at the Data Repository for the University of Minnesota (DRUM).}

\suppdata{Supporting Information is provided. This includes data about code evaluations and additional data not presented in the main text.}

% Set the IOP-specific numerical reference style
\bibliographystyle{iopart-num}

% Specify your Zotero-imported .bib file (omit the .bib extension here)
\bibliography{references}

\end{document}